\documentclass[11pt,twoside]{article}
\usepackage[utf8]{inputenc}
\usepackage[a4paper,twoside,bindingoffset=1cm,head=15pt,body={17cm,25cm},top=2.5cm]{geometry}
\usepackage{amsmath,amssymb}
\usepackage{graphicx}
\usepackage{color}
\usepackage{cite}
\usepackage{ifthen}
\usepackage{titlesec}
\usepackage{fancyhdr}

\titleformat{\section}{\normalfont\bfseries}{\thesection}{1em}{}
\titlespacing*{\section}{0pt}{*1.5}{1ex}

\newcounter{authorcounter}

\newcommand{\wavesauthorlist}{}
\newcommand{\wavesaddresslist}{}
\newcommand{\wavesemail}{}
\newcommand{\wavesfootnotes}{}
\newcommand{\wavesauthorpre}{}

\def\theNumberTest#1{%
  \if\relax\detokenize\expandafter{\romannumeral-0#1}\relax
    true%
  \else
    false%
  \fi
}

\newcommand{\wavesspeaker}[3][noemail]{%
    \ifthenelse{\value{authorcounter} > 1}{%
      \renewcommand{\wavesauthorpre}{, }%
    }{%
      \renewcommand{\wavesauthorpre}{}%
    }%
    \ifthenelse{\equal{#1}{noemail}}{%
      \renewcommand{\wavesfootnotes}{}
    }{%
      \renewcommand{\wavesemail}{$^\ast$Email: #1}%
      \renewcommand{\wavesfootnotes}{, \ast}
    }%
    \ifthenelse{\equal{\theNumberTest{#3}}{true}}{%
      \edef\wavesauthorlist{\wavesauthorlist%
        \wavesauthorpre{}\underline{#2}$^{#3\wavesfootnotes%
        }$%
      }%
    }{%
      \edef\wavesauthorlist{\wavesauthorlist%
        \wavesauthorpre\underline{#2}$^{\arabic{authorcounter}\wavesfootnotes%
        }$%
      }
      \edef\wavesaddresslist{\wavesaddresslist%
        \par%
        $^{\arabic{authorcounter}}$#3%
      }%
      \stepcounter{authorcounter}%
    }%
  \ignorespaces
}

\newcommand{\wavesauthor}[3][noemail]{%
    \ifthenelse{\value{authorcounter} > 1}{%
      \renewcommand{\wavesauthorpre}{, }%
    }{%
      \renewcommand{\wavesauthorpre}{}%
    }%
    \ifthenelse{\equal{#1}{noemail}}{%
      \renewcommand{\wavesfootnotes}{}
    }{%
      \renewcommand{\wavesemail}{$^\ast$Email: #1}%
      \renewcommand{\wavesfootnotes}{, \ast}
    }%
    \ifthenelse{\equal{\theNumberTest{#3}}{true}}{%
      \edef\wavesauthorlist{\wavesauthorlist%
        \wavesauthorpre{}#2$^{#3\wavesfootnotes%
        }$%
      }%
    }{%
      \edef\wavesauthorlist{\wavesauthorlist%
         \wavesauthorpre{}#2$^{\arabic{authorcounter}\wavesfootnotes%
         }$%
      }
      \edef\wavesaddresslist{\wavesaddresslist%
        \par%
        $^{\arabic{authorcounter}}$#3%
      }%
      \stepcounter{authorcounter}%
    }%
  \ignorespaces
}

\newenvironment{wavespaper}[3]{%
  \renewcommand{\wavesauthorlist}{}%
  \renewcommand{\wavesemail}{}%
  \setcounter{authorcounter}{1}%
     #2
  \twocolumn[
    \begin{center}
     \bfseries
     #1
     \bigskip

     \wavesauthorlist
     \mdseries
     \smallskip

     \wavesaddresslist
     \smallskip
 
     \wavesemail

    \end{center}%
  ]

}{%
}

\fancypagestyle{plain}{%
\fancyhead{}
}

\usepackage[utf8]{inputenc}
\usepackage{graphicx}
\usepackage{latexsym}
\begin{document}
\begin{wavespaper}{%
Non-Equilibrium Magnetosonic Wave Motion
}{%
  \wavesspeaker[ellermeier@fkp.tu-darmstadt.de]{Wolfgang F. Ellermeier}{Department of Physics, TU Darmstadt, Darmstadt, Germany}
}{%
Xavier Antoine, Ricardo Weder}
\noindent\textbf{thermodynamic relaxation, magnetohydrodynamics,piston analogy}
\small
\abstract{In ideal compressible hydrodynamics there is an isomorphism between spatially one-dimensional unstea- dy and two-dimensional steady supersonic flow called piston analogy [7]. This notice shows that this is also true for non-equilibrium magnetosonic flow under alignment of undisturbed flow and magnetic field in case of steady flow. 
An example for two generic problems, i.e. the signal problem of radiation into a half space and steady flow along a kinked wall bounding a half space, is given.}
\section{Basic Equations}
The MHD equations of  motion for a thermodynamically relaxing fluid of infinite electrical conductivity and a single non-equilibrium process
comprise the equation of mass, the momentum equation, the induction equation, the energy balance equation, the relaxation equation and the 
thermodynamic equations of state. They are written, in respective order, see [5]:
\begin{eqnarray}
\dot \varrho +\nabla \cdot (\varrho \vec V)=0,\\
\varrho (\dot {\vec V}+(\vec V\cdot \nabla) \vec V)=-\nabla p+\nonumber \\
\frac{1}{\mu_0}(\nabla \times\vec B)\times \vec B,\\
\dot {\vec B}=\nabla \times (\vec V\times \vec B),\\
\varrho (\dot h+\vec V\cdot \nabla h)-\dot p-\vec V\cdot \nabla p=0,\\
\dot \xi +\vec V\cdot \nabla \xi =-\frac{\xi -\tilde \xi (p,\varrho )}{\tau (p,\varrho ,\xi )},\\
h=\hat h(p,\varrho,\xi).
\end{eqnarray}
The field variables are mass density $\varrho$, flow velocity $\vec V$, pressure $p$, magnetic field $\vec B$,and specific enthalpy $h$. The thermodynamic state function is $\hat h$, the relaxation time $\tau$ is a stricltly positive function of state [3]. The inner variable $\xi$ with its equilibrium value $\tilde \xi$ 
describes either the degree of reaction in a single chemical reaction (ionization-recombination) or a vibrational non-equilibrium process in the plasma [4].
Next, linearization of the above equations is performed for small perturbations of a state of thermodynamic equilibrium (state of rest or uniform flow) in the presence of a constant background magnetic field $\vec B_0$:
\begin{eqnarray}
&&\dot \varrho+\varrho_0 \nabla \cdot \vec V=0,\\
&&\varrho_0 \dot{\vec V}=-\nabla p+\frac{1}{\mu_0}(\nabla \times \vec b)\times \vec B_0,\\
&&\dot{\vec b}=\nabla \times (\vec V \times \vec B_0),\\
&&\varrho_0 h-p=0,\\
&&\tau_0\dot \xi+\xi=\tilde \xi_{p0}p+\tilde \xi_{\varrho0}\varrho,\\
&&h=\hat h_{p0}p+\hat h_{\varrho0}\varrho+\hat h_{\xi0}\xi.
\end{eqnarray}
The relaxation time is $\tau_0:=\tau(p_0,\varrho_0,\xi_0)$ and $\varrho(\vec x,t)$, $\vec b(\vec x,t),...$ are the perturbations of
mass density, magnetic field strength etc.. 
Omitting details the linearised flow field vector equation 
\begin{eqnarray}
(T_0\,\frac{\partial}{\partial t}+\frac{a_{f0}^2}{a_{e0}^2})\ddot{\vec V}-
a_{f0}^2(T_0\,\frac{\partial}{\partial t}+1)\nabla \nabla \cdot \vec V \,=\nonumber \\
(T_0\frac{\partial}{\partial t}+\frac{a_{f0}^2}{a_{e0}^2})\nabla \times \nabla \times(\vec V \times \frac{ \vec B_0}{\sqrt{\mu_0 \varrho_0}})\times\frac{
\vec B_0}{\sqrt{\mu_0\varrho_0}}
\end{eqnarray}
with the definitions
\begin{align}
a_{f0}^2:=\frac{\hat h_{\varrho_0}}{\varrho_0^{-1}-\hat h_{p_0}},\,\,a_{e0}^2:=\frac{\hat h_{\varrho_0}+\hat h_{\xi _0}\tilde \xi _{\varrho_0}}
{\varrho_0^{-1}-\hat h_{p_0}-\hat h_{\xi_0}\tilde\xi_{p_0}},\nonumber\\
T_0:=\tau(p_0,\varrho_0,\xi_0)\frac{\hat h_{\varrho_0}}
{\hat h_{\varrho_0}+\hat h_{\xi_0}\tilde \xi_{\varrho_0}},\nonumber
\end{align}
is derived. The constants $a_{f0},a_{e0}$ are the frozen and equilibrium sound speeds, respectively. 
Thermodynamic considerations (i.e. stability of the rest state and positive entropy production rate (see [3,5])) imply
\begin{equation}
\tau_0>0,\,T_0>0,\,a_{f0}>a_{e0}.
\end{equation}
Obviously for $\nabla \cdot \vec V=0$ thermodynamic relaxation plays no roll at all; then the governing equation is
\begin{eqnarray}
\ddot{\vec V}=
\nabla \times \nabla \times(\vec V \times \frac{ \vec B_0}{\sqrt{\mu_0 \varrho_0}})\times\frac{
\vec B_0}{\sqrt{\mu_0\varrho_0}}
\end{eqnarray}
and describes Alfven shear waves.
Plane magnetosonic relaxing flow is specified by
\begin{eqnarray}
&&\vec V=u(x,y,t)\vec e_x+v(x,y,t)\vec e_y,\\
&&\vec b=a(x,y,t)\vec e_x+b(x,y,t)\vec e_y,\\
&&\vec B_0=A_0\vec e_x+B_0 \vec e_y, 
\end{eqnarray}
with the background magnetic field in the $x,y$-plane at an angle $\,\arctan(B_0/A_0)\,$
between $\vec B_0$ and the $x$-axis, orthonormal unit vectors $\vec e_{x},\vec e_{y}$ along the coordinate axes are used. 
\section{Unsteady one-dimensional flow}
\begin{figure}[htbp]
	\centering
		\includegraphics[width=0.50\textwidth]{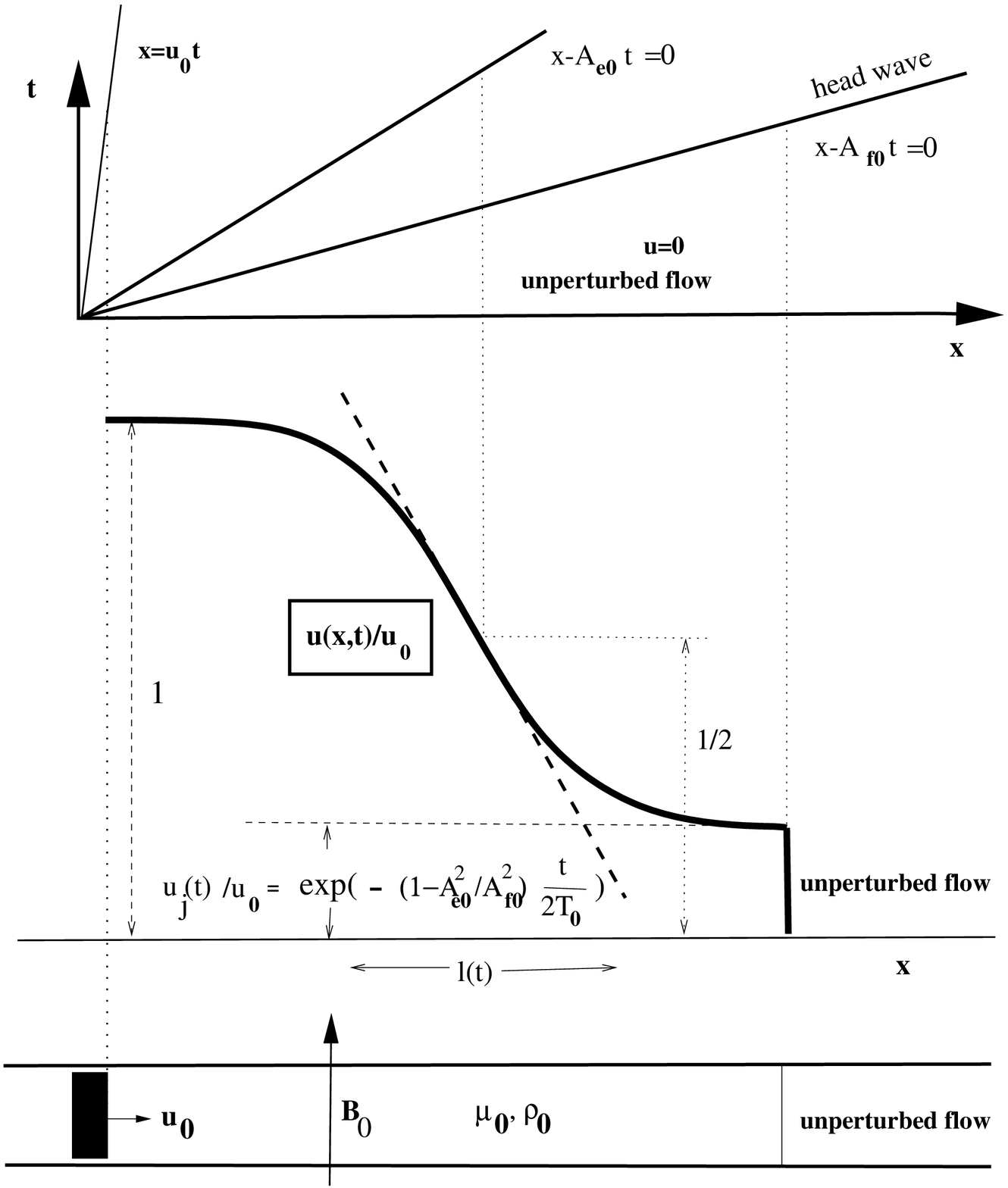}
		 \caption{Partially dispersed wave in relaxing MHD flow, unsteady velocity-wave moving from
      left to right with decaying front discontinuity invading thermodynamic equilibrium
        state to the right of discontinuity; the front height decays exponentially in time according to eq. (42) as it moves to the right. Both, s-shape and front discontinuity travel faster with increase of the background magnetic field $B_0$ regardless of upwards or downwards orientation. The flow is realized in a shock tube with the piston at one end suddenly put to constant speed $u_0$ and the magnetic field transverse to the tube axis.}
	\label{fig:voll}
\end{figure}
For $A_0=0,v\equiv 0,a\equiv 0,b=b(x,t),u=u(x,t)$, i.e. $\partial/\partial y\equiv 0$ one derives from eq. (13),
\begin{equation}
T_0(\ddot u-A_{f0}^2u_{xx})\,\dot{}+
\ddot u-A_{e0}^2u_{xx}
=0,
\end{equation}
where definitions according to
\begin{eqnarray}
&&A_{f_0,e_0}^2:=\frac{B_0^2}{\mu_0\varrho_0}+a_{f_0,e_0}^2,
\end{eqnarray}
are employed. The specification of initial and boundary conditions completes the mathematical formulation:
\small
\begin{eqnarray}
u(0,t)=u_0H(t),\,\,u(x,0)=0,
\end{eqnarray}
\small
with $H(t)$ denoting the Heaviside jump.
There are three wave speeds , $B_0/ \sqrt{\mu_0\varrho_0},
a_{f0},a_{e0}$ involved now rather than just two in the case without
magnetic field or just one in absence of both. 
The above equation is written in a form to bring out the analogy with the relaxation equation of
ordinary gasdynamics [1,2], to which it reduces for vanishing $B_0$:
\begin{equation}
T_0(\ddot u-a_{f0}^2u_{xx})\,\dot{}+
\ddot u-a_{e0}^2u_{xx}
=0,
\end{equation}
In [1,2] solutions of the last equation are presented for the piston driven tube with generating
the flow by imposing a constant piston velocity $u_0$ for $t>0$, so that a compression wave
is generated. The formal method to substantiate this picture quantitively 
is to time-Laplace transform eqs. (19,21) and taking account of causality.  Since the evolution equation (19) with magnetic field $B_0$ is identical in  form to eq. (22) one can reinterpret the solutions of the magneto-free case accordingly. The results are, for the front
discontintinuity $u_j(t)$ (Fig. 1):
\begin{equation}
u_j(t)=u_0\,e^{-\frac{a_{f0}^2-a_{e0}^2}{A_{f0}^2}\,\frac {t}{2T_0}},
\end{equation}
the non-oscillatory $s$-shaped relaxation zone behind the front broadens its width $l(t)$ according to
\begin{equation}
l(t)= \sqrt{2tT_0(a_{f0}^2-a_{e0}^2)},
\end{equation}
and the turning point of the $s$-shape ($u=u_0/2$) lags behind the front at a constant rate $A_{f0}-A_{e0}$; the asymptotic representation of the solution for $t\,\gg \,2T_0A_{f0}^2/(A_{f0}^2-A_{e0}^2)$,
takes the form
\begin{align}
u(x,t)\sim \frac{1}{2}u_0\left(1-\phi(\frac{x-A_{e0}t}{l(t)})\right),
\,\phi(a):=\frac{2}{\sqrt{\pi}}
\int_0^ae^{-b^2}db.
\end{align}
Note that $l(t)$ is a positive quantity since due to the last inequality (14).
The influence of the magnetic field can be summarized as follows:\newline
-the front discontinuity's speed $A_{f0}$
increases with the field strength and is also larger than the front speed in the absence of magnetic fields,
\newline - the jump height of the front discontinuity decreases more slowly with increasing
    $\mid B_0\mid $,i.e. the fastest decay is achieved in the absence of magnetic background fields,\newline
-the speed gap between discontinuity and relaxation zone decreases with increasing magnetic field and is largest for vanishing magnetic field,\newline
-for $\mid B_0 \mid/\sqrt{\mu_0\varrho_0}\gg a_{f0}(>a_{e0})$, i.e. for sufficiently strong magnetic field, the piston problem
is equivalent to the case without relaxation producing a constant $v$ distribution between piston and
front discontinuity of constant height $u_0$ traveling at the speed $A_{f0}$
into the undisturbed gas. The four statements apply regardless of the sign of $B_0$.
\newline
The high and low frequency asymptotics approximation of eq. (19) produces a telegraph type equation of the form
\begin{align}
\ddot u-A_{f0}^2u_{xx}+\frac{1}{T_0}\frac{a_{f0}^2-a_{e0}^2}{A_{f0}^2}\dot u=0,\nonumber
\end{align}
in the first and a Kelvin-Voigt type equation
\begin{align}
\ddot u-A_{e0}^2u_{xx}=T_0(a_{f0}^2-a_{e0}^2)\dot u_{xx},\nonumber 
\end{align}
in the second case, the factor $T_0(a_{f0}^2-a_{e0}^2)$ representing a positive coefficient of diffusion, which is independent of the background magnetic field strength. For unidirectional wave motion one finds 
\begin{align}
\dot u+A_{f0}u_x+\frac{a_{f0}^2-a_{e0}^2}{2T_0A_{f0}^2}u=0, \nonumber\\
\dot u+A_{e0}u_x-\frac{T_0}{2}(a_{f0}^2-a_{e0}^2)u_{xx}=0 \nonumber ,
\end{align}
or, in comoving frames at speeds $A_{f0}$ and $A_{e0}$, resp., the signal evolution is governed by
\begin{align}
\dot u+\frac{a_{f0}^2-a_{e0}^2}{2T_0A_{f0}^2}u=0,\\
\dot u-\frac{T_0}{2}(a_{f0}^2-a_{e0}^2)u_{xx}=0.
\end{align}
Whereas the telegraph equation is of hyperbolic type and does admit discontinuous solutions the Kelvin-Voigt type approximation is strictly parabolic in terms of mathematical classification [6]. The partially dispersed wave of Fig.1 exhibits both features as can be found by inspecting eqs. (23,25); it turns out that eq. (25) simply is a solution of the diffusion equation (27) and eq.(23) solves the unidirectional telegraph equation (26)on $x=A_{f0}t$.
Thus the hyperbolic part of eq. (19) is associated with the behaviour of the signal along the wave head separating disturbed and undisturbed regions in the $x,t$-plane, whereas the Kelvin-Voigt part of eq. (19) desribes the behaviour sufficiently far downstream of the head wave (Fig. 2). The complete solution of the signal problem for the regime between the head wave and the $t$-axis cannot, however,  be expressed in terms of elementary functions.
\section{Steady two-dimensional flow for aligned fields}
\begin{figure}[ht]
	\centering
		\includegraphics[width=0.50\textwidth]{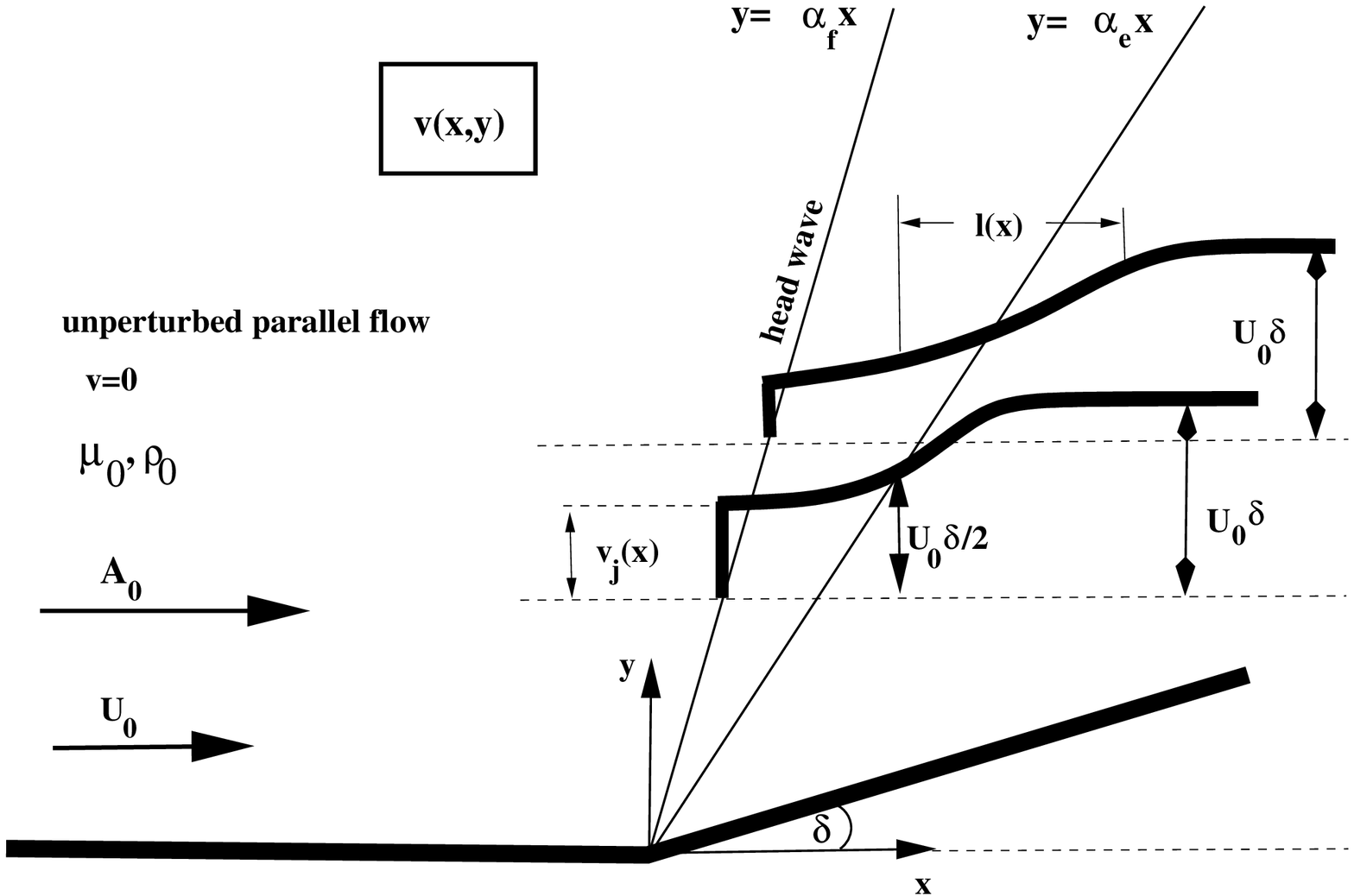}
		\caption{Distribution of $v(x,y)$ in steady  flow of a relaxing fluid along kinked wall bounding a half space with aligned magnetic background field. Upstream of head wave there is undisturbed parallel flow, in the far field away from the wall the flow approaches uniform flow with $v\,\rightarrow\, U_0\delta,u\,\rightarrow \,U_0$.}
	\label{fig:wedgerelaxmhd}
\end{figure}
For $B_0=0,u\rightarrow U_0+u(x,y),v=v(x,y),p=p(x,y), \\ b=b(x,y),a\rightarrow A_0+a(x,y), \partial/\partial t=U_0 \partial/\partial x$
and definitions for the constant coefficients, i.e. the frozen and equilibrium Mach numbers $M_{f,e}$ and the Alven-Mach number $M_a$, i.e.
\begin{eqnarray}
\alpha_{f,e}^2:=\frac{M_a^2-1+M_{f,e}^2}{(M_a^2-1)(M_{f,e}^2-1)},\\
M_{f,e}:=\frac{U_0}{a_{f0,e0}},\,\,\,M_a:=\frac {U_0}{A_0/\sqrt{\mu_0\varrho_0}},
\end{eqnarray}
one obtains the following governing equation for $v(x,y)$ from eq. (13):
\begin{eqnarray}
T_0U_0\,\frac{M_f^2-1}{M_e^2-1}\left(v_{xx}-\alpha_f^2 v_{yy}\right)_x+v_{xx}-\alpha_e^2 v_{yy}=0.
\end{eqnarray}
It may be worthwhile to indicate, both, the incompressible and the field free  versions of the above equation, i.e.
\begin{align}
v_{xx}+v_{yy}=0, \nonumber\\
T_0U_0[(M_f^2-1)v_{xx}-v_{yy}]_x+(M_e^2-1)v_{xx}-v_{yy}=0,\nonumber
\end{align}
respectively. The first indicating simply potential flow without the magnetic field affecting the flow at all, and the second being
known from the literature[1,2].
Furthermore, the case of compressible flow without relaxation, as described in [8] is also recovered:
\begin{align}
v_{xx}-\frac{M_a^2-1+M^2}{(M_a^2-1)(M^2-1)}v_{yy}=0,
\end{align}
with $M$ denoting the ordinary Mach number based on the isentropic sound speed of non-relaxing material.
\small
The boundary conditions to eq. (30) (with $H(x)$ Heaviside jump) are
\begin{align}
v(x,0)=H(x) U_0 \delta.
\end{align}
For the case $M_f\,>\,1$, so that $M_e\,>\,1$ as well and  $M_a\,>\,1$ so that the flow velocity is supersonic w.r.t. all three wave speeds $a_{f0,e0},A_0/\sqrt{\varrho_0\mu_0}$ there is equivalence between the 
signal problem according to eqs. (19,21) and the steady flow problem by eqs. (30,32) with for instance $u_0\hat {=}\delta U_0, u\hat = v$; for the piston analogy to hold true the wall has to turn into the stream so that the material is compressed rather than expanded, i.e. $0<\delta(<<1) $.
As a consequence one can immediately describe the solution for the kinked wall problem by reinterpretation of the partially dispersed
wave problem and vice versa. The pertinent equations analogous to eqs. (26,27) of the unsteady problem, reexpressing $\alpha_{e,f}$ in terms of the Mach numbers and the length $T_0U_0$,  are
\begin{align}
v_x+\frac{v}{L}=0,\,L:=2T_0U_0\frac{M_f^2-1}{M_e^2-M_f^2}\frac{M_a^2-1+M_f^2}{M_a^2}\\
v_x-L'\,v_{yy}=0,\,L':=\frac{T_0U_0}{2}\frac{M_a^2}{M_a^2-1}\frac{M_e^2-M_f^2}{(M_e^2-1)^2}.
\end{align}
Both coefficients $L,L'$ in the above equations are positive under the assumptions $M_f,M_a$ each larger than unity.
A qualitative representation of the result is depicted in Fig. 2 showing  a jump in $v$, i.e. $v_j(x)=U_0 \delta \exp{(-x/L)}$,
decaying away from the wall downstream of the wall kink along the head wave. The $s$-shaped spreading zone behind the discontinuity  is governed by
\begin{align}
v(x,y) \sim \frac{U_0 \delta}{2}\left( 1-\phi \left( \frac{y-\alpha_e x}{l(x)}\right)\right),
\end{align}
with $\phi$ denoting the error function as in the unsteady case and 
\begin{align}
l(x):=\sqrt{2T_0U_0\frac{M_a^2}{M_a^2-1}\frac{M_e^2-M_f^2}{(M_e^2-1)^2}\,x}.
\end{align}


\end{wavespaper}

\end{document}